\documentclass[conference]{IEEEtran}
\IEEEoverridecommandlockouts

\usepackage{listings}
\newenvironment{algo2lst}[1][htb]
  {
   \begin{algorithm}[#1]%
}{\end{algorithm}}

\DeclareFixedFont{\ttb}{T1}{txtt}{bx}{n}{8.5} 
\DeclareFixedFont{\ttm}{T1}{txtt}{m}{n}{8.5}  

\usepackage{color}
\definecolor{deepblue}{rgb}{0,0,0.5}
\definecolor{deepred}{rgb}{0.6,0,0}
\definecolor{deepgreen}{rgb}{0,0.5,0}

\newcommand\pythonstyle{\lstset{
    language=Python,
    basicstyle=\ttm,
    aboveskip=10pt,
    belowskip=10pt,
    otherkeywords={self},             
    keywordstyle=\ttb\color{deepblue},
    emph={MyClass,__init__},          
    emphstyle=\ttb\color{deepred},    
    stringstyle=\color{deepgreen},
    numbers=none,
    frame=tb,                         
    showstringspaces=false            %
}}

\lstnewenvironment{python}[1][]
{
    \pythonstyle
    \lstset{#1}
}{}

\usepackage{color}
\usepackage{fancyvrb}
\usepackage{fvextra}
\usepackage[ruled, linesnumbered, noend]{algorithm2e}
\SetAlFnt{\footnotesize}
\SetAlCapFnt{\footnotesize}
\SetAlCapNameFnt{\footnotesize}
\SetAlCapHSkip{0pt}
\IncMargin{-\parindent}
\SetInd{0.75em}{0.1em}

\usepackage{caption} 
\usepackage{multirow} 

\DefineVerbatimEnvironment{Highlighting}{Verbatim}{commandchars=\\\{\}, breaklines=true}

\newcommand{\devito}{\href{https://github.com/devitocodes/devito}{Devito} }

\usepackage[unicode=true]{hyperref}
\hypersetup{breaklinks=true,
            pdfauthor={},
            pdftitle={Temporal blocking of finite-difference stencil operators with sparse ``off-the-grid'' sources},
            colorlinks=true,
            citecolor=black,
            urlcolor=blue,
            linkcolor=black,
            pdfborder={0 0 0}}
\usepackage{cite}
\usepackage{amsmath,amssymb,amsfonts}
\usepackage{graphicx}
\usepackage{textcomp}
\usepackage[table]{xcolor}    
\usepackage{longtable,booktabs}
\usepackage{caption}
\usepackage{subcaption}
\def\BibTeX{{\rm B\kern-.05em{\sc i\kern-.025em b}\kern-.08em
    T\kern-.1667em\lower.7ex\hbox{E}\kern-.125emX}}

\begin{document}

\title{Temporal blocking of finite-difference stencil operators with sparse ``off-the-grid" sources}

\makeatletter
\newcommand{\linebreakand}{%
  \end{@IEEEauthorhalign}
  \hfill\mbox{}\par
  \mbox{}\hfill\begin{@IEEEauthorhalign}
}
\makeatother

\author{
	\IEEEauthorblockN{George Bisbas}
	\IEEEauthorblockA{\textit{Imperial College London}\\
		London, UK \\
		g.bisbas18@imperial.ac.uk}
	\and
	\IEEEauthorblockN{Fabio Luporini}
	\IEEEauthorblockA{\textit{Devito Codes}\\
		London, UK \\
		fabio@devitocodes.com}
	\and
	\IEEEauthorblockN{Mathias Louboutin}
	\IEEEauthorblockA{\textit{Georgia Institute of Technology}\\
		Atlanta, GA \\
		mlouboutin3@gatech.edu}
	\linebreakand 
	\IEEEauthorblockN{Rhodri Nelson}
	\IEEEauthorblockA{\textit{Imperial College London}\\
		London, UK \\
		rnelson@imperial.ac.uk}
	\and
	\IEEEauthorblockN{Gerard J. Gorman}
	\IEEEauthorblockA{\textit{Imperial College London}\\
		London, UK \\
		g.gorman@imperial.ac.uk}
	\and
	\IEEEauthorblockN{Paul H.J. Kelly}
	\IEEEauthorblockA{\textit{Imperial College London}\\
		London, UK \\
		p.kelly@imperial.ac.uk}
}


\maketitle

\begin{abstract}\label{abstract}
	Stencil kernels dominate a range of scientific applications, including seismic and medical imaging, image processing, and neural networks. Temporal blocking is a performance optimization that aims to reduce the required memory bandwidth of stencil computations by re-using data from the cache for multiple time steps. It has already been shown to be beneficial for this class of algorithms. However, applying temporal blocking to practical applications' stencils remains challenging. These computations often consist of sparsely located operators not aligned with the computational grid (``off-the-grid"). Our work is motivated by modelling problems in which source injections result in wavefields that must then be measured at receivers by interpolation from the grided wavefield. The resulting data dependencies make the adoption of temporal blocking much more challenging. We propose a methodology to inspect these data dependencies and reorder the computation, leading to performance gains in stencil codes where temporal blocking has not been applicable. We implement this novel scheme in the Devito domain-specific compiler toolchain. Devito implements a domain-specific language embedded in Python to generate optimized partial differential equation solvers using the finite-difference method from high-level symbolic problem definitions. We evaluate our scheme using isotropic acoustic, anisotropic acoustic, and isotropic elastic wave propagators of industrial significance. After auto-tuning, performance evaluation shows that this enables substantial performance improvement through temporal blocking over highly-optimized vectorized spatially-blocked code of up to 1.6x.
\end{abstract}

\begin{IEEEkeywords}
	temporal blocking, stencil computations, code generation, partial differential equations, seismic imaging, domain-specific languages, wave-propagation
\end{IEEEkeywords}

\section{Introduction}

Stencils are commonly encountered in scientific applications such as image processing \cite{Ragan-Kelley2018}, convolutional neural networks, weather forecasting \cite{Stella}, computational fluid dynamics, seismic \cite{LuporiniTOMS, Louboutin2019}, and medical imaging \cite{guasch2020full}. We present a scheme that enables the application of temporal blocking, a common technique to enhance cache-locality \cite{Wonnacott2000, Jin2001,Wonnacott2004AchievingSL}, to a class of finite-difference (FD) \cite{LeVeque2007}, \cite{smith1985finite} stencil kernels where this is challenging. This class consists of additional sparsely-located in the computational grid (off-the-grid) operators. Typical stencil kernels are computational patterns that are usually functions of the nearest neighboring point values. In a more general context, a stencil defines the iterative computation of an element in an $n$-dimensional spatial grid at time $t$ as a function of neighboring grid elements (space dependencies) at time $t-1, \dots, t-k$ (time dependencies). A typical stencil update in a scientific simulation has a 3-dimensional spatial iteration space and 1-dimensional temporal iteration space. Figure \ref{fig:1d3pt_update} illustrates a 1D stencil and its flow dependences. Each point is updated using values from the previous timestep and the right and left neighbors. Arrows illustrate the flow dependencies. Halo points (grey) are used to extend the computational domain by the stencil radius size. Wider stencils in 3D and their respective data dependencies are illustrated in Figure \ref{fig:halos-stencils}.

\begin{figure}[!ht]
    \centering
    \includegraphics[width=0.30\textwidth]{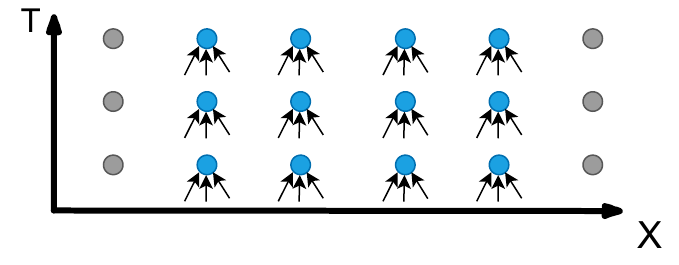}
    \caption{A 1D-3pt Jacobi stencil update. Arrows show the data flow dependencies, grey points indicate the read-only halo area.}
    \label{fig:1d3pt_update}
\end{figure}

\begin{figure}[!htbp]
    \centering
    \captionsetup[subfigure]{labelformat=empty}
    \subfloat[]{\includegraphics[width=0.300\hsize]{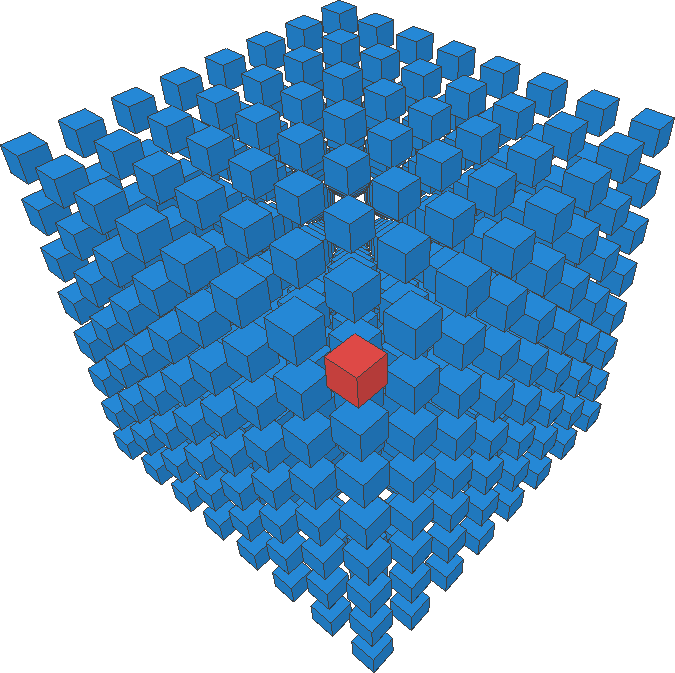}}
    \subfloat[]{\includegraphics[width=0.300\hsize]{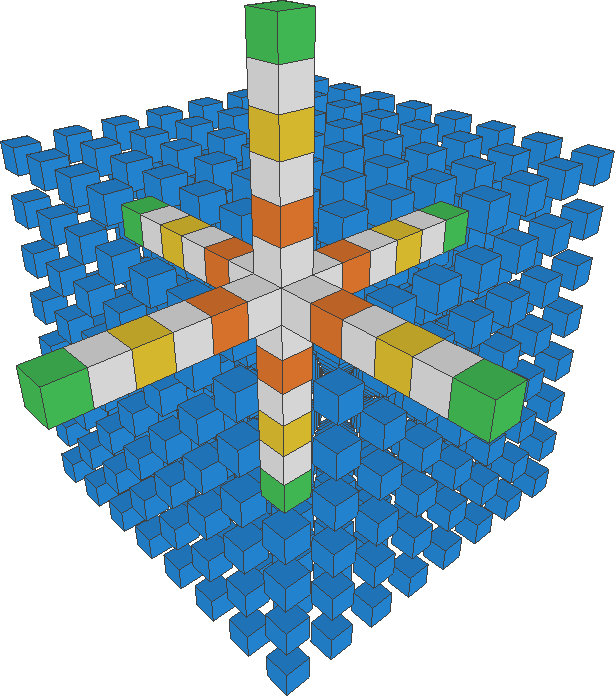}}
    \subfloat[]{\includegraphics[width=0.300\hsize]{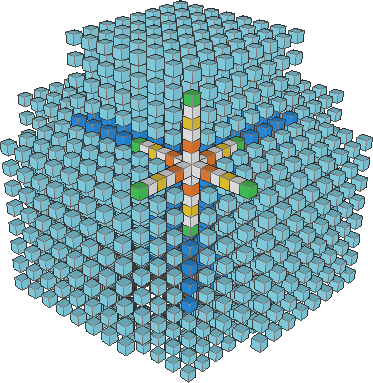}}
    \caption{A 6th-order ($O(1,8)$) 3D-19pt stencil update. A point (red) at the edge of a block (blue) depends on a four-deep halo of neighbouring points which extends outside the block.}\label{fig:halos-stencils}
\end{figure}

In addition to data dependencies of the kind illustrated in Figure 1, applications such as seismic wave-modeling carry additional dependencies owing to the interpolation of data not directly associated with grid points into the model (e.g., source injection). These positions, that are not aligned with the grid points are sparsely-distributed off-the-grid \cite{Yan2015, poon2019degrees} positions as shown in Figure \ref{fig:off_the_grid_src}. They may also include receivers that interpolate neighboring values to take measurements, as shown in Figure \ref{fig:off_the_grid_rec}. Sources and receivers are sets of sparsely-distributed off-the-grid points. We iterate over these sparsely-located sets through indirections applying their effect to the grid points after iterating the 3D grid for stencil updates for each timestep. A loop nest structure illustrating this computation pattern is shown in Listing \ref{lst:typical_src_stencil}, where $nt$ is the number of time steps; $nx$, $ny$, $nz$ are the number of grid points along the $x$, $y$, $z$ dimensions respectively, $A(t, x, y, z)$ is the stencil kernel update and $so$ is the space discretisation order. The {\tt src} is of size $nt \times len(sources)$ holding the wavefield for each timestep for every source where $sources$ is the structure holding the information for modeling source injection. Variable {\tt np} accounts for the number of points affected from a source, and $f$ is the function defining the type of interpolation (e.g., bilinear, trilinear). An example of bilinear interpolation is shown in Figure \ref{fig:off-the-grid},  where 4 points are affected in 2D space. The {\tt sources} set shown in Listing \ref{lst:typical_src_stencil} provides the sparse off-the-grid coordinates for the injection. We iterate this set of coordinates that determine the affected neighboring points. The wave amplitude is scattered to these affected points.


\begin{figure}[!htbp]
    \centering
    \begin{subfigure}[b]{0.24\textwidth}
        \includegraphics[width=0.75\textwidth]{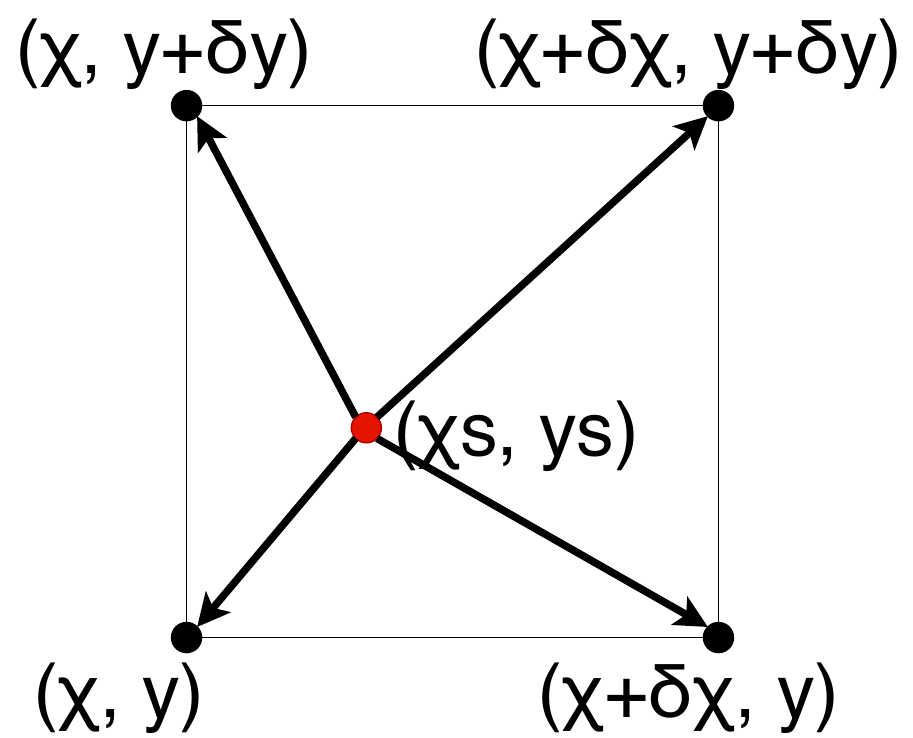}
        \caption{An off-the-grid source injects values to neighboring grid points.}
        \label{fig:off_the_grid_src}
    \end{subfigure}
    \begin{subfigure}[b]{0.24\textwidth}
        \includegraphics[width=0.75\textwidth]{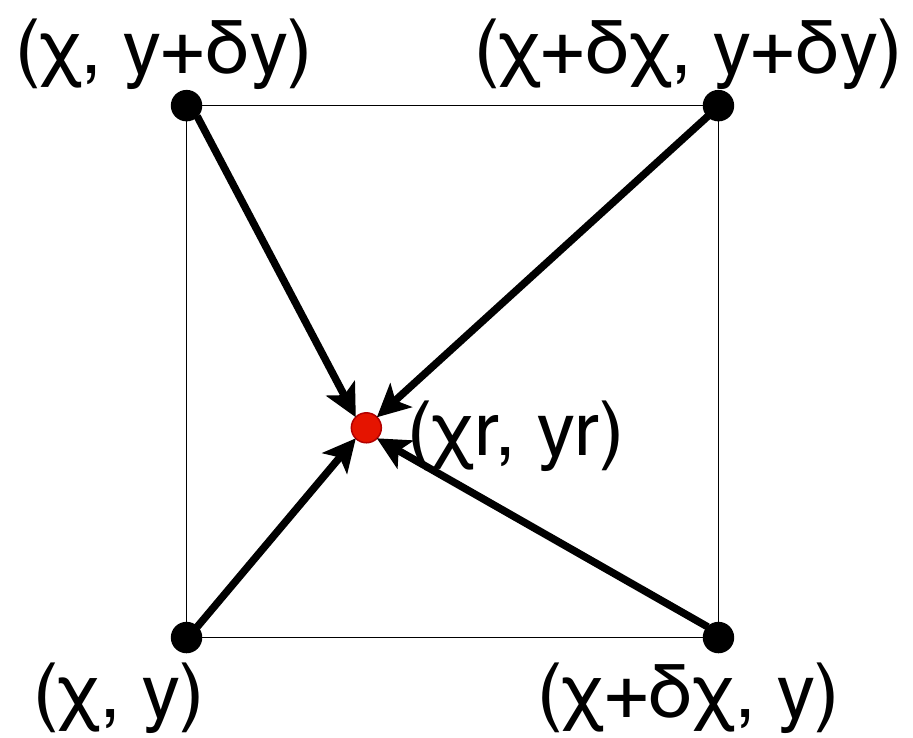}
        \caption{An off-the-grid receiver interpolates values from neighboring grid points.}
        \label{fig:off_the_grid_rec}
    \end{subfigure}
    \hfill
    \caption{A source injection and a receiver measurement interpolation at off-the-grid positions in a 2-D FD-grid. We assume linear interpolation.}\label{fig:off-the-grid}
\end{figure}

\begin{algo2lst}[!htbp]
    \For{ t = {\tt 1} \textbf{to} {\tt nt}}{
        \For{ x = {\tt 1} \textbf{to} {\tt nx}}{
            \For{ y = {\tt 1} \textbf{to} {\tt ny}}{
                \For{ z = {\tt 1} \textbf{to} {\tt nz}}{
                    $A(t, x, y, z) \equiv$ u[t, x, y, z] = u[t-1, x, y, z] + 
                    $\sum_{r=1}^{r=so/2}w_r\Big($ \
                    u[t-1, x - r, y, z] + u[t-1, x + r, y, z] +
                    u[t-1, x, y - r, z] + u[t-1, x, y + r, z] +
                    u[t-1, x, y, z - r] + u[t-1, x, y, z + r] \Big);
                    }
                }
            }
		\ForEach(\tcp*[h]{For every source}){s \textbf{in} {\tt sources}}{
			\For(\tcp*[h]{Get the points affected}){ i = {\tt 1} \textbf{to} {\tt np}}{
				xs, ys, zs = map(s, i)\ \tcp*[h]{through indirection} \\
					u[t, xs, ys, zs] + = $f(src(t, s)$) \tcp*[h]{add their impact on the field}
			}
        }
	}
    \caption{A typical time-stepping loop nest structure for a stencil update with source injection. This stencil has one temporal and three spatial dimensions.}\label{lst:typical_src_stencil}
\end{algo2lst}

\subsection{Problem overview: a running example}\label{overview}

Space blocking \cite{Wolfe1989, Wolf1991} can be applied in computational patterns similar to Listing \ref{lst:typical_src_stencil}, but applying temporal blocking is challenging as we illustrate with a 1-D example in Figure \ref{fig:space_blocking_violation}. Red diamonds indicate off-the-grid coordinates where sparse operators are applied. Sparse operators are applied after a time iteration for the whole domain is finished. Consequently, separating the FD grid into blocks does not violate any data dependencies.

\begin{figure}[!ht]
    \centering
    \begin{subfigure}[b]{0.48\textwidth}
        \centering
        \includegraphics[width=0.66\textwidth]{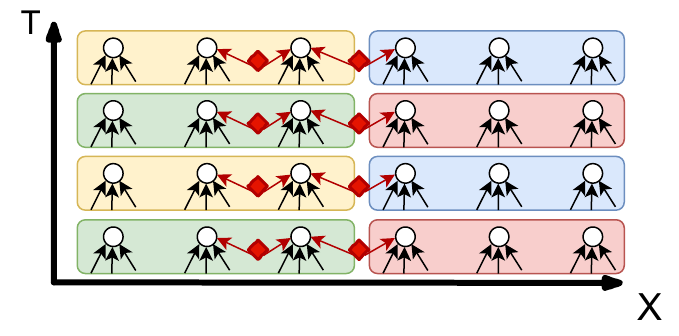}
        \caption{Rectangular space blocking. All grid points can be updated in parallel at a specific time-step. Sparse operators fit within space blocking as their effect is imposed after all points have been updated.}
        \label{fig:space_blocking}
    \end{subfigure}
    \hfill
    \begin{subfigure}[b]{0.48\textwidth}
        \centering
        \includegraphics[width=0.66\textwidth]{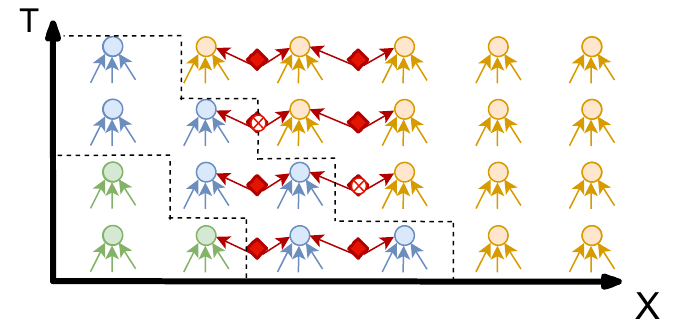}
        \caption{Skewed/Wave-front temporal blocking. Grid points are updated in waves. During a wave-front update, we compute grid point values for multiple timesteps. Applying sparse operators in the space boundary may lead to erroneous updates since source injection may precede the stencil update for a particular timestep. We have a data dependency violation.}
        \label{fig:violation}
    \end{subfigure}
    \hfill
    \caption{Sparse operators do not violate data dependencies in space blocking, Fig. \ref{fig:space_blocking} in contrast to temporal blocking, Fig. \ref{fig:violation}.}
    \label{fig:space_blocking_violation}
\end{figure}

In contrast to space blocking, temporal blocking cannot be applied. When a sparse operator is located at an off-the-grid position, among points that belong to different space blocks, data dependencies are violated, thus yielding incorrect results. The violation occurs because updates in space may pause for a particular time-step, and computation will proceed in time rather than space. Consequently, a sparse operator update may be computed, and points that have not yet been updated through the stencil kernel updates may be affected. Similarly, a point may be erroneously updated due to a forward move in time but may miss injection from a neighboring off-the-grid operator due to space-time block constraint. Data dependencies are violated, and similar violations are raised in other variants of temporal blocking, such as wave-front temporal blocking \cite{Yount2016}, \cite{Wellein2009} diamond temporal blocking \cite{Bertolacci2015}, \cite{Malas2014} and others. We aim to overcome this limitation through our contributions in this paper.

Because the set of sources is sparse, the loops generated in Listing \ref{lst:typical_src_stencil} by modelling source injection consist of non-affine accesses as illustrated in Listing \ref{lst:typical_src_stencil}. While polyhedral tools such as PLUTO \cite{Bondhugula2008, Bondhugula2013}, Polly \cite{Grosser2012PollyP}, Loopy \cite{kloeckner_loopy_2014}, and CLooG \cite{Bastoul2004} manage to deal with the first uniform stencil update, they are not capable of dealing with the non-affine nature of the source injection loop nests.

The sparse off-the-grid nature of the source operator combined with the non-affine nature of our loop structure illustrated in Listing \ref{lst:typical_src_stencil} is blocking the application of time-tiling as described in subsection \ref{overview}. The methodology presented in Section \ref{methodology_and_implementation} aims to overcome this limitation. 

\subsection{Related work}

\subsubsection{Improving stencil performance}\label{improving_performance} stencils offer good parallelisation opportunities, ranging from Instruction Level Parallelism (ILP), SIMD register-level parallelism (SSE, AVX) to shared-memory (OpenMP, OpenACC) and task parallelism. Furthermore, distributed-memory parallelism is often employed.

In most applications of interest, stencil kernels have low operational intensity, having few floating-point operations per byte of data accessed, and are therefore memory bandwidth-bound. Considerable effort has been put into caches for improving stencil performance. Rescheduling the order of computations towards increased cached memory reuse can increase throughput by alleviating memory bandwidth boundedness issues.

\paragraph{Spatial cache blocking}\label{cache_blocking} as illustrated in \cite{Wolfe1989, Wolf1991, Ramanujam1992, Xue1997tiling, Frigo2005, Frigo2006, Datta2008, Tang2011, Frigo2012} FD grids can easily be decomposed into tiles and benefit from cache blocking. The same idea is applicable to unstructured grids, though this requires more sophisticated algorithms to create efficient schedules \cite{luporini2019automated}. Space tiling techniques have also been implemented for execution on GPUs. Related work includes automated split tiling with trapezoids \cite{Grosser2013}, hybrid hexagonal tiling \cite{Grosser2014} and automated HPC GPU code \cite{Schafer2011}, \cite{Holewinski2012}, \cite{Rawat2018} as well as hybrid spatial/temporal blocking on FPGAs \cite{Zohouri2018}.

\paragraph{Temporal cache blocking}\label{temporal_cache_blocking} extending cache reuse in the time dimension led to the development of temporal blocking algorithms. To further reduce cache misses, we utilize computed values in a block to update values in the next timestep where possible. While one or more timesteps for a given block's values are stored in the cache, we start computing the next time step for this block, not depending on the requirement to compute all the blocks of a given grid for previous timesteps. Spatial and temporal reuse are often fused into hybrid models (equidistant locality) to harness the advantages of both methods \cite{Zohouri2018}. Plenty of research has been conducted in designing and evaluating temporal blocking schemes ranging from simple skewing \cite{Wolfe1986, Wonnacott2000, Jin2001, Wonnacott2004AchievingSL, Wellein2009, Strzodka2011} and wave-front \cite{Yount2016} to more sophisticated such as diamond \cite{Bertolacci2015, Bandishti2012, Malas2014, Muranushi2015OptimalTB, levchenko2017, Akbudak2020AsynchronousCF}. The technique presented in this paper enables such schedules to be used in applications with off-the-grid operators. While narrow stencil kernels exhibit good temporal locality, temporal blocking gains decrease when space-order increases. Higher space order problems limit temporal locality as more space updates are required to update one value in time. 

\subsubsection{Domain Specific Languages}\label{dsls} improving performance is essential but usually comes with the price of error-prone hand-optimization. Finding ways to automate HPC code generation led to the birth of several domain-specific Languages (DSLs) like Devito \cite{LuporiniTOMS, Louboutin2019}, which is used in this paper. Several DSL/compiler frameworks are working towards the automated generation of PDE solvers, such as FEniCS \cite{FEniCSProject2015}, and Firedrake \cite{Rathgeber2017}. Halide \cite{Ragan-Kelley2018}, implemented as an internal DSL in C++. OpenSBLI \cite{Jacobs2017OpenSBLIAF} is another framework that generated C code from Python-based high-level abstractions targetting equations written in Einstein notation. Halide is a language targeting code generation for digital image processing featuring memory locality and vectorized computation optimizations and ported to multi-core CPUs and GPUs.

Other automatic code-generation frameworks are OPS \cite{Reguly2014} for GPU code generation and YASK \cite{Yount2015}, a DSL to create high-performance FD-stencil code. Lift \cite{Hagedorn2018}, achieves performance portability on parallel accelerators by combining high-level functional data-parallel language with rewrite rules which encode algorithmic and hardware-specific optimization choices. Stella \cite{Stella} and GridTools \cite{Thaler2019} are DSLs embedded in C++, focusing on weather and climate HPC simulations.

\subsection{Contributions}\label{contribution}

Our contributions are:

\begin{itemize}
    \item We propose a scheme that precomputes the off-the-grid sparse operators' effect, allowing to reorder the computations for FD wave propagators, thus enabling the application of temporal blocking to stencil codes consisting of sparse operators such as source injection and measurement interpolation. Our scheme is cost-efficient, adding a negligible overhead compared to the measured gains.
    \item We implement the algorithm directly on top of the Devito DSL, harnessing the power of automated code generation, thus providing a pathway to express any similar operator in a form that exploits the benefits of time tiling with only minimal coding effort.
    \item We evaluate our scheme using 3D stencils encountered in wave propagation applications (isotropic acoustic, isotropic elastic, and anisotropic acoustic (TTI)), each having different memory and compute requirements.
    \item We achieve performance gains ranging from 15\% to 60\% for space order 4 and 8 for isotropic acoustic and elastic and anisotropic acoustic as well as 5\% to 10\% gains for elastic and TTI cases at space order 12. 
\end{itemize}

Our work is mainly motivated by the need to enable and automate these optimizations for a class of seismic and medical imaging problems. Characteristic examples of such applications include full-waveform inversion (FWI) \cite{virieux2009overview} and reverse time migration (RTM) \cite{baysal1983reverse}. As future work, we aim to deliver these optimizations as a fully automated workflow.

The rest of the paper is organized as follows: Section \ref{methodology_and_implementation} presents the approach followed to solve our problem. Section \ref{kernels} introduces the wave-propagation kernels to be evaluated, and Section \ref{evaluation} presents an experimental evaluation of the applicability and impact of the approach. Finally, in Section \ref{conclusion}, we discuss and summarise our work and briefly refer to our plans for future work.

\section{Methodology and implementation}\label{methodology_and_implementation}

In this section, we describe our approach that enables temporal blocking for wave-propagators with sparse operators. We describe the individual steps and present the details of our implementation. The whole precomputation workflow benefits from the power of the \devito DSL \cite{LuporiniTOMS, Louboutin2019} to automatically generate code and the data structures required by our scheme in its DSL. Afterward, we manually transform the generated loops to implement wave-front temporal blocking (WTB) \cite{Wonnacott2004AchievingSL, Ramanujam1991, Lamport1974}, a representative temporal blocking schedule.

\subsection{Source injection precomputation}\label{precomputation}

The source injection modeling consists of the following parameters: the number of sources, their coordinates, and their wavelet time-series. This data is enough to precompute their effect on an empty grid. We assume that the sources' coordinates are constant across our models' time-domain though this may not always be the case. However, Devito's API can support the moving sources' case, and our algorithm is independent of it.

\subsubsection{Iterate sources' coordinates and store indices of affected points} initially, we iterate over each source and inject to an empty grid for one timestep, assuming the wavefield is not zero at the first timestep. If the wavefield is zero at the first timestep, we may inject for more timesteps. Our experiments use wavefields with non-zero values at the first timesteps. The pseudocode is illustrated in \ref{lst:inject_one}. We use {\devito} to automatically generate code for this step. Our scheme is independent of the injection and interpolation type (e.g., non-linear injection). Then, we store the non-zero grid point coordinates.

\begin{algo2lst}[!htbp]
	\For{ t = 1 \textbf{to} {\tt 2}}{
        \ForEach{s \textbf{in} {\tt sources}}{
			\For{ i = {\tt 1} \textbf{to} {\tt np}}{
				xs, ys, zs = map(s, i)\;
					u[t, xs, ys, zs] + = $f(src(t, s)$)
			}
        }
	}
	\caption{Source injection over an empty grid. No PDE stencil update is happening.}\label{lst:inject_one}
\end{algo2lst}

\subsubsection{Generate sparse binary mask and unique IDs} using the nonzero indices, we populate two arrays. The first array (Fig. \ref{fig:mask1}) is a binary integer mask of our grid with $1$s at indices where $u$ is nonzero. Ones are shown as filled bullet circles, with a green background, Fig. \ref{fig:mask1}. The second one (Fig.\ref{fig:ids}) has the same shape and is populated with unique ascending values for each unique point affected. It is quite common to encounter points being affected by more than one source. Figures show an x-y plane (z-slice) of the 3D grid.

\begin{figure}[!htbp]
	\centering
	\begin{subfigure}[b]{0.24\textwidth}
		\includegraphics[width=0.90\textwidth]{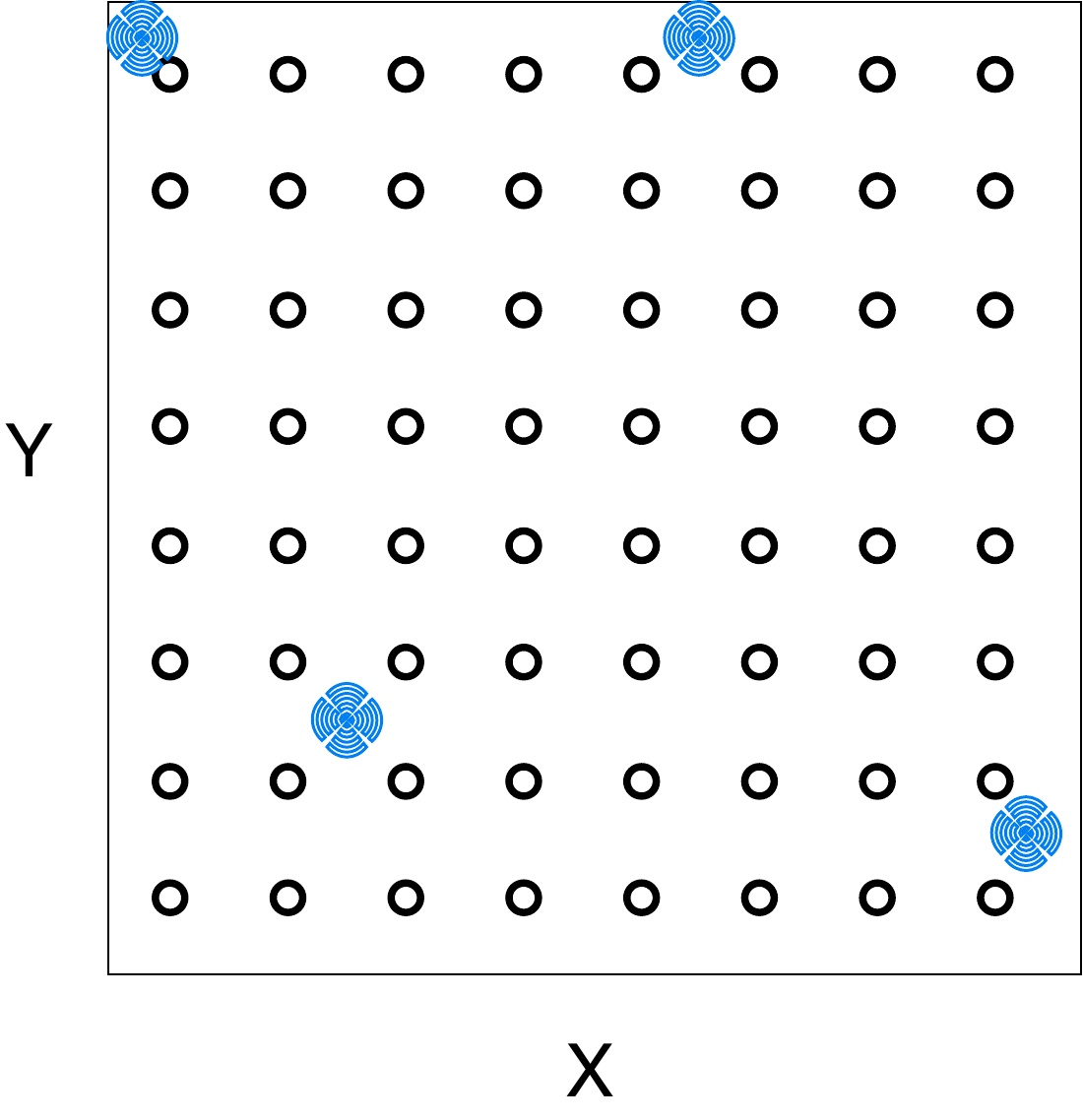}
		\caption{Sparsely located sources at off-the-grid positions.}
		\label{fig:nonaligned}
	\end{subfigure}
	\begin{subfigure}[b]{0.24\textwidth}
		\includegraphics[width=0.90\textwidth]{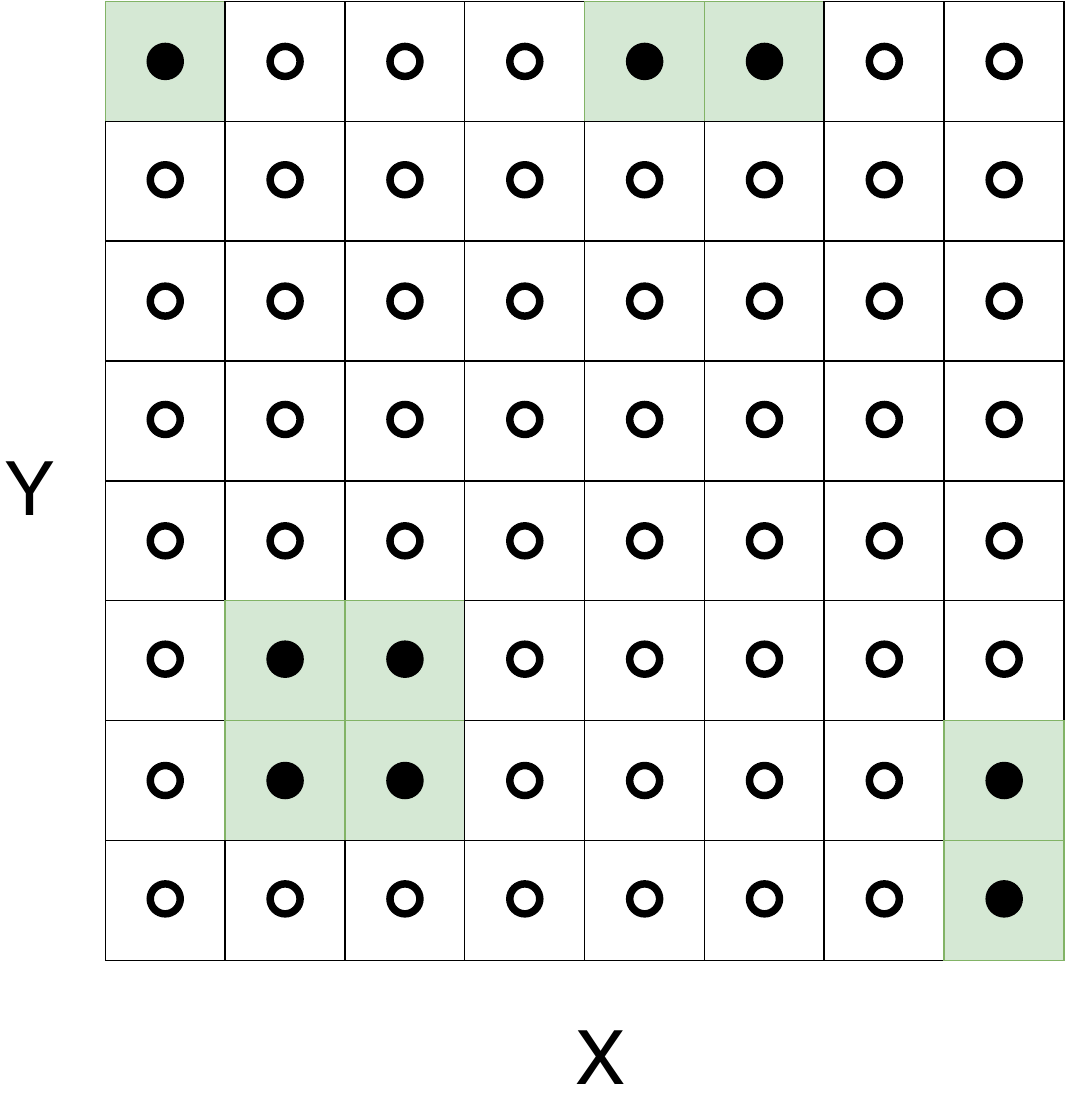}
		\caption{Identify unique points affected ({\tt SM}). \hfill}
		\label{fig:mask1}
	\end{subfigure}
	\hfill
	\begin{subfigure}[b]{0.24\textwidth}
		\includegraphics[width=0.90\textwidth]{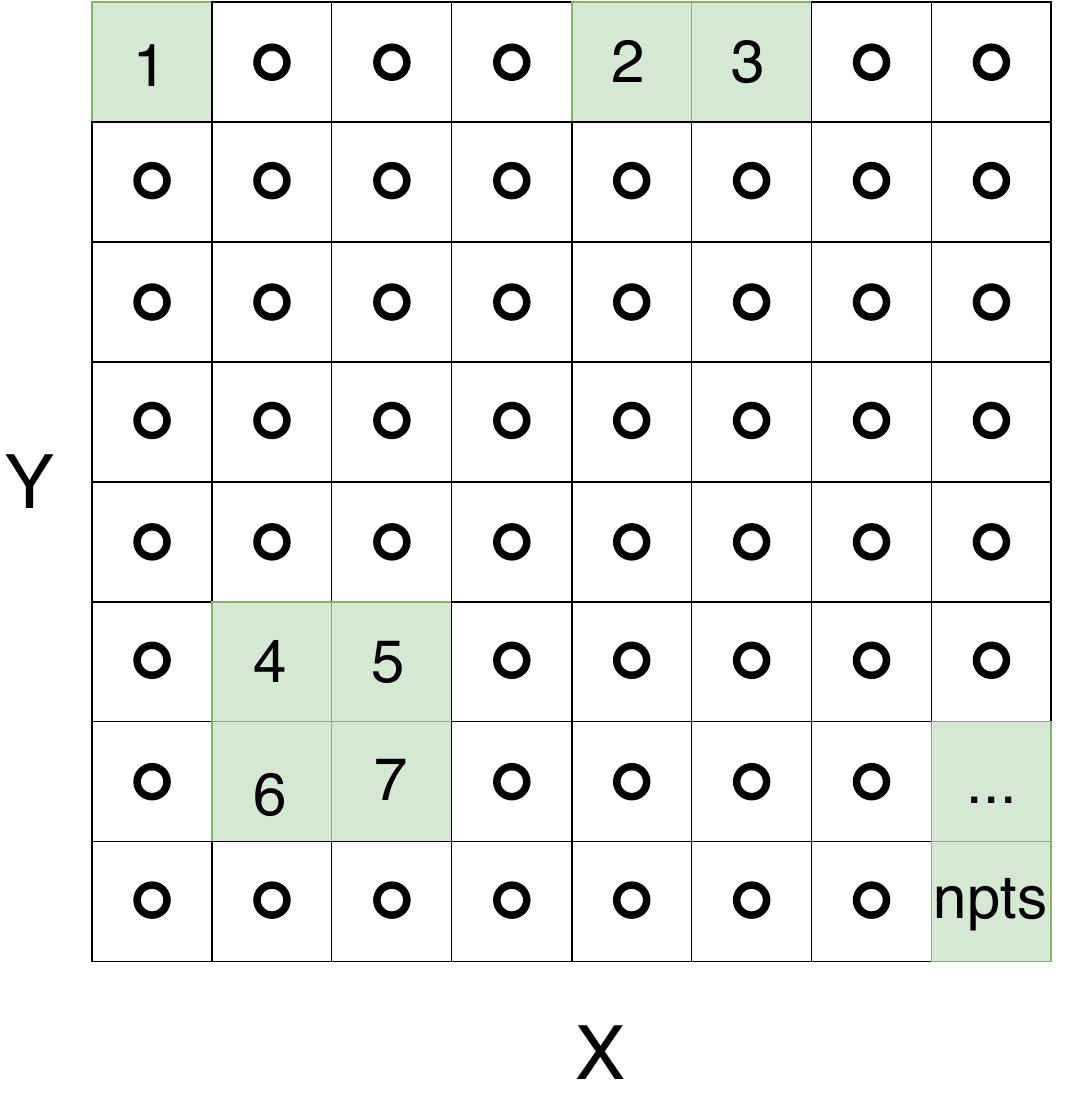}
		\caption{Assign a unique ID to each of the {\tt npts} affected points ({\tt SID}).}
		\label{fig:ids}
	\end{subfigure}
	\begin{subfigure}[b]{0.24\textwidth}
		\includegraphics[width=0.90\textwidth]{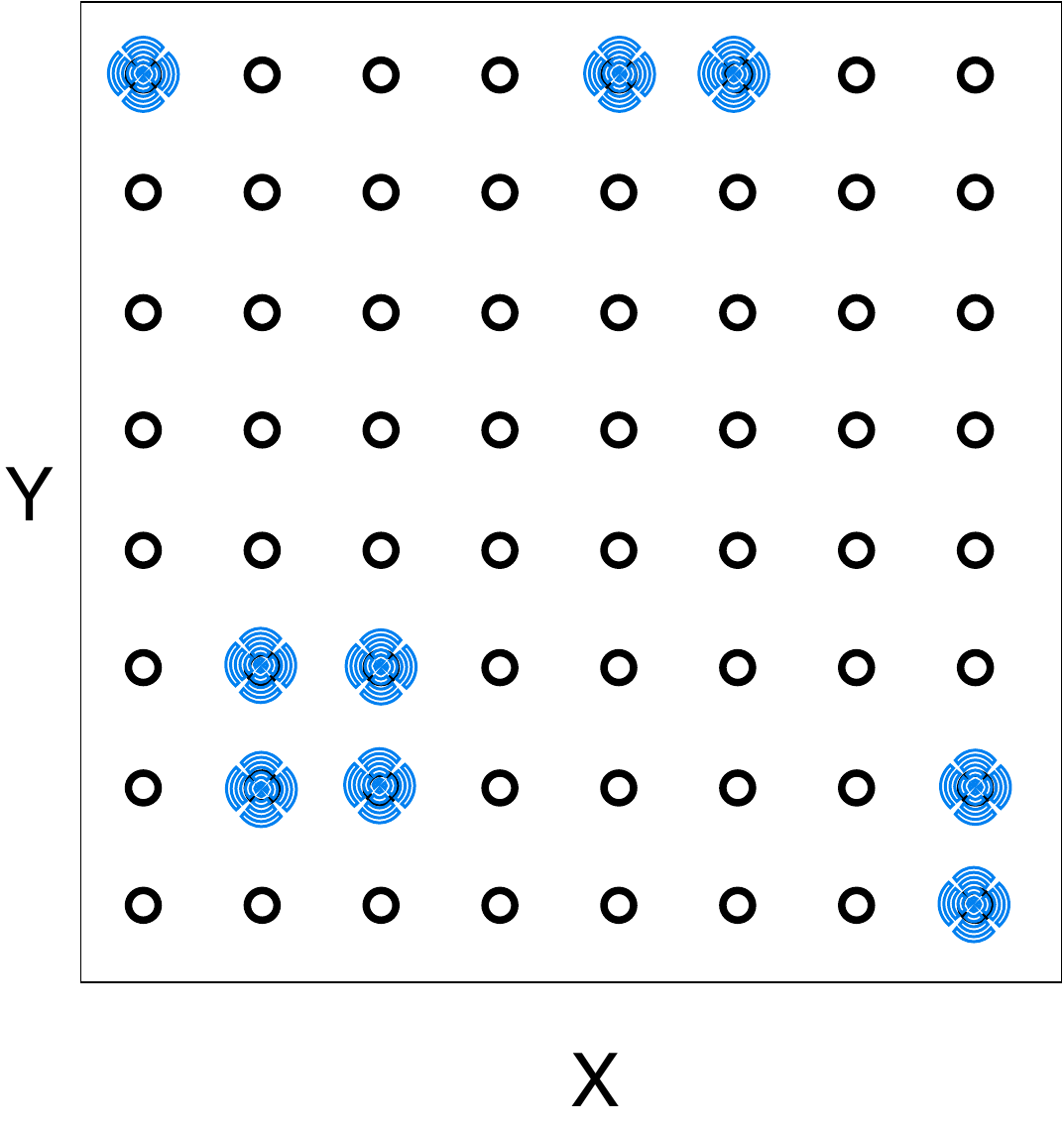}
		\caption{{\tt npts} sources are now aligned with grid point coordinates.}
		\label{fig:aligned}
	\end{subfigure}
	\caption{Illustration of the four steps through which source impact is aligned to the computational grid. The figures show an x-y plane slice of the 3D grid.} \label{steps}
\end{figure}

\subsubsection{Decompose wavefields} knowing the unique positions affected and their coordinates, we now use Devito's source injection mechanism to decompose the off-the-grid positioned wavefields to grid-aligned point wavefields. Using the {\tt SID} structure, we perform an indirection and decomposition of the sources' wavefields to per-affected-point wavefields. The pseudocode for that workflow is presented in Listing \ref{lst:inject_two}. {\tt src\_dcmp} now replaces {\tt src} in our source injection computations. Instead of having sources at off-the-grid positions (Fig.\ref{fig:nonaligned}), we now have decomposed, aligned to the grid point sources (Fig.\ref{fig:aligned}).

\begin{algo2lst}[!htbp]
	\For{ t = {\tt 1} \textbf{to} nt}{
		\ForEach{{\tt s} \textbf{in} {\tt sources}}{
			\For{ i = {\tt 1} \textbf{to} {\tt np}}{
				xs, ys, zs = map(s, i)\;
				src\_dcmp[t, SID[xs, ys, zs]] + = $f(src(t, s)$;
			}
		}
	}
	\caption{Decomposing the source injection wavefields.}\label{lst:inject_two} 
\end{algo2lst}

\subsubsection{Fuse iteration spaces} using the aligned structure {\tt src\_dcmp}, we can now fuse the source injection loop inside the kernel update iteration space. There is no {\tt sources} loop as sparse data can be expressed in 3D coordinates. We fuse the source injection loop at the same loop level as the stencil update {\tt z} loop. The source mask {\tt SM} acts as a binary mask and is used to add (if 1) or not (if 0) the source impact while {\tt SID} is used to access the impact values indirectly as we iterate over the grid dimensions. The resulting loop structure is illustrated in the following pseudocode in Listing \ref{lst:new_struct} and also offers SIMD vectorization opportunities over the {\tt z2} loop.

\begin{algo2lst}[!htbp]
	\For{ t = {\tt 1} \textbf{to} {\tt nt}}{
		\For{ x = {\tt 1} \textbf{to} {\tt nx}}{
			\For{ y = {\tt 1} \textbf{to} {\tt ny}}{
				\For{ z = {\tt 1} \textbf{to} {\tt nz}}{
					$A(t, x, y, z, s)$;
					}
				\For{ z2 = {\tt 1} \textbf{to} {\tt nz}}{
					u[t, x, y, z2] + = SM[x, y, z2] * src\_dcmp[t, SID[x, y, z2]];
				}
			}
		}
	}
	\caption{Stencil kernel update with fused source injection.}\label{lst:new_struct}
\end{algo2lst}

\subsubsection{Reducing the iteration space size} the 3D structures that are used to iterate through sources ({\tt SM} and {\tt SID}) in the {\tt z2} loop are, in the general case, massively sparse. Multiplications by zero are dominant. Only the necessary iterations in {\tt z} dimension need to be performed to alleviate this issue. We aggregate nonzero occurrences along the z-axis of {\tt SM} recording them in a structure named {\tt nnz\_mask}. We reduce the size of {\tt SID} cutting off z-slices where all elements are zero. For naming convention, we use {\tt Sp\_SID} for the new structure. These structures reduce the iteration space size of {\tt z2} to perform the necessary computation. Pseudocode for the new structure is illustrated in Listing \ref{lst:new_struct_reduced}. The opportunity to reduce the iteration space generally applies to the majority of problems in seismic. However, we show that benefits are not limited in cases where the reduction is small (see subsection \ref{corner_cases}).

\begin{figure}[!htbp]
	\centering
	\begin{subfigure}[b]{0.16\textwidth}
		\includegraphics[width=\textwidth]{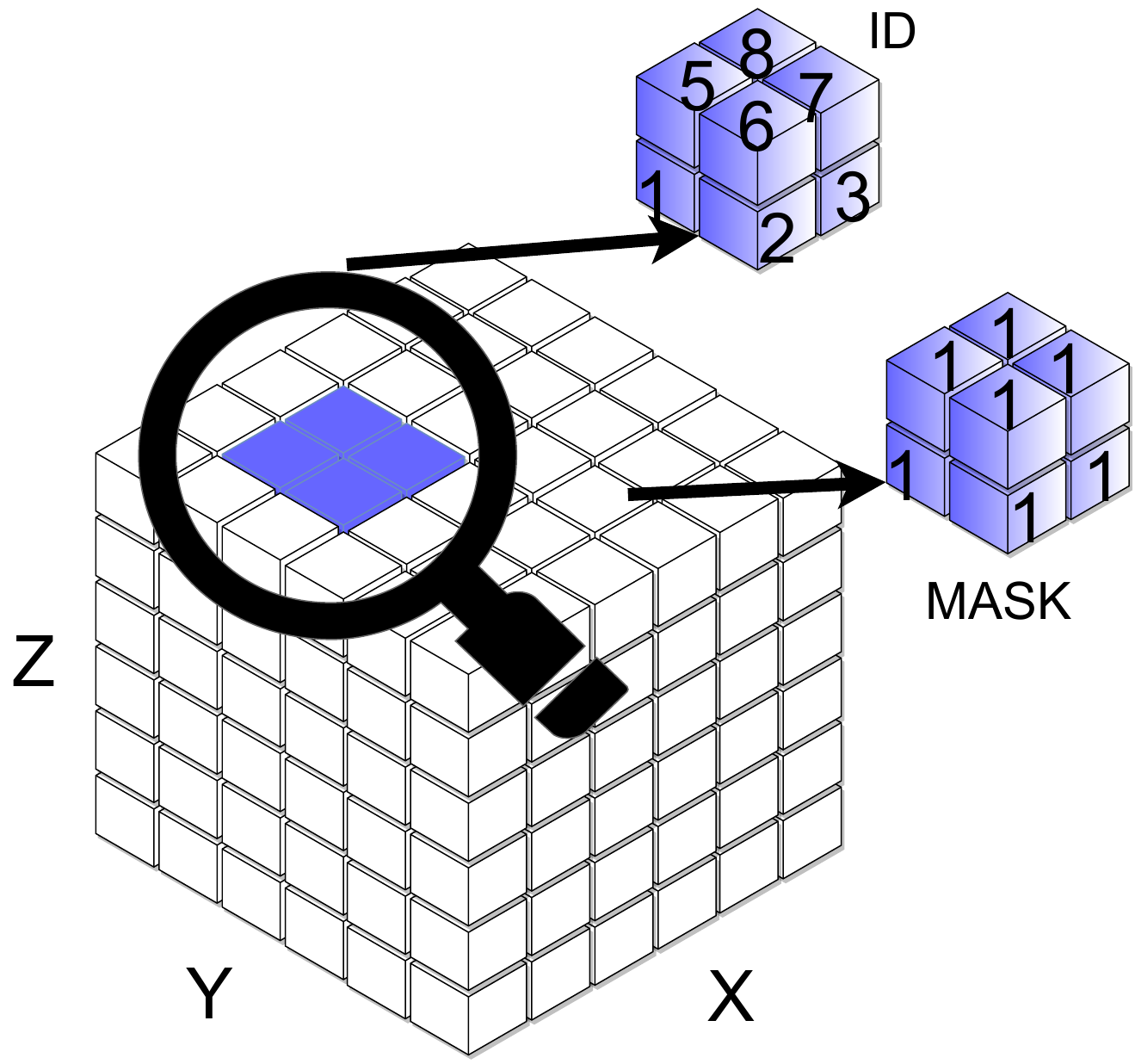}
		\caption{{\tt SID} and {\tt SM} are very sparse in the general case.} \hfill
		\label{fig:mask_n_id}
	\end{subfigure}
	\begin{subfigure}[b]{0.15\textwidth}
		\includegraphics[width=\textwidth]{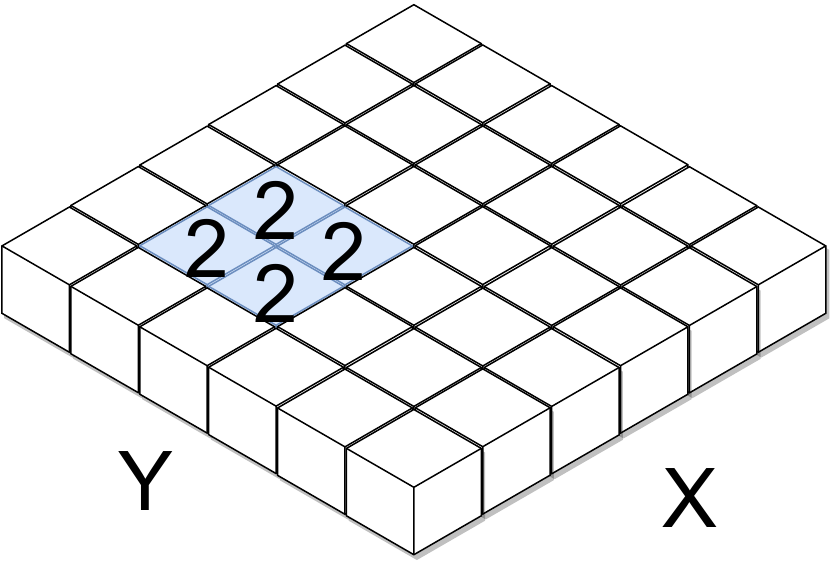}
		\caption{{\tt nnz\_mask} Aggregating non-zero values along z-axis.}
		\label{fig:aggregation}
	\end{subfigure}
	\begin{subfigure}[b]{0.15\textwidth}
		\includegraphics[width=\textwidth]{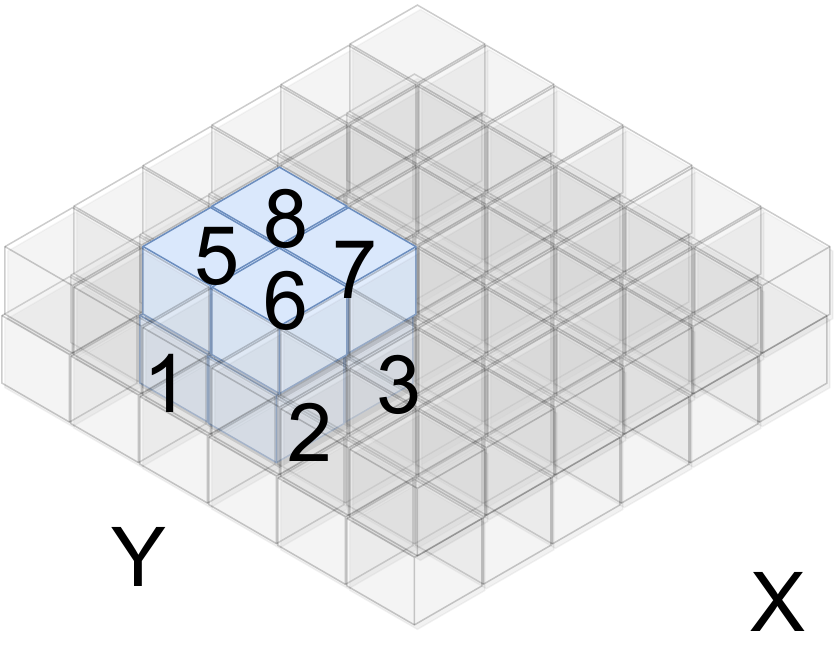}
		\caption{{\tt Sp\_SID}, a reduced size {\tt SID}. } \hfill \hfill
		\label{fig:sparse_ids}
	\end{subfigure}
	\caption{We aggregate nonzero occurrences along the z-axis, keeping count of them. The size of {\tt SID} is reduced by cutting off z-slices where all elements are zero.} \label{sparse_save}
\end{figure}

\begin{algo2lst}[!htbp]
	\For{ t = {\tt 1} \textbf{to} {\tt nt}}{
		\For{ x = {\tt 1} \textbf{to} {\tt nx}}{
			\For{ y = {\tt 1} \textbf{to} {\tt ny}}{
				\For{ z = {\tt 1} \textbf{to} {\tt nz}}{
					$A(t, x, y, z, s)$;
					}
				\For{ z2 = {\tt 1} \textbf{to} {\tt nnz\_mask[x][y]}}{
					$I(t, x, y, z) \equiv$\{
					zind = Sp\_SID[x, y, z2]\;
					u[t, x, y, z2] += src\_dcmp[t, SID[x, y, zind]];
					\}
					  }
				}
			}
		}
	\caption{Stencil kernel update with reduced size iteration space for source injection.}\label{lst:new_struct_reduced}
\end{algo2lst}

Finally, we present a methodology that aligns the source injection impact to the grid points, thus enabling the application of TB to stencil operators with sparse off-the-grid points.

\subsection{Applying wave-front temporal blocking}\label{wavefront_temporal}

We present the loop transformations to apply WTB to the loop structure in Listing \ref{lst:new_struct_reduced}. In temporal blocking, we extend space blocking so that multiple timesteps are evaluated in a subset of the overall problem domain. In WTB space-time, wave-fronts traverse our domain computing grid point values. For naming convention, as used in \cite{Yount2016}, we are going to use the term ``block" for spatial-only grouping and ``tile" when multiple temporal updates are allowed. Fig. \ref{fig:txy_wavefront} shows grid point updates as they happen in WTB. The green point is updated using orange values. This stencil kernel has a radius of size two. Thus a margin of 2 points is required to preserve data dependencies. Only two timesteps are kept in memory (for time order one problems), so the green value substitutes the yellow one in the buffer. The stencil radius affects the wave-front angle (the ratio of spatial indices needed to update one point in the next time step). The angle gets steeper with a higher stencil radius. WTB can also be applied to staggered grids. In this case, two or more grids may be updated, often having inter-dependencies \cite{Yount2016}. It is then necessary to shift the wave-front angle (allow more margin to preserve dependencies) by an amount, depending on the stencil radius of data dependencies in each loop as shown in Figure \ref{fig:stag_seq_wavefront}.

\begin{figure}[!htpb]
	\centering
	\includegraphics[width=0.4\textwidth]{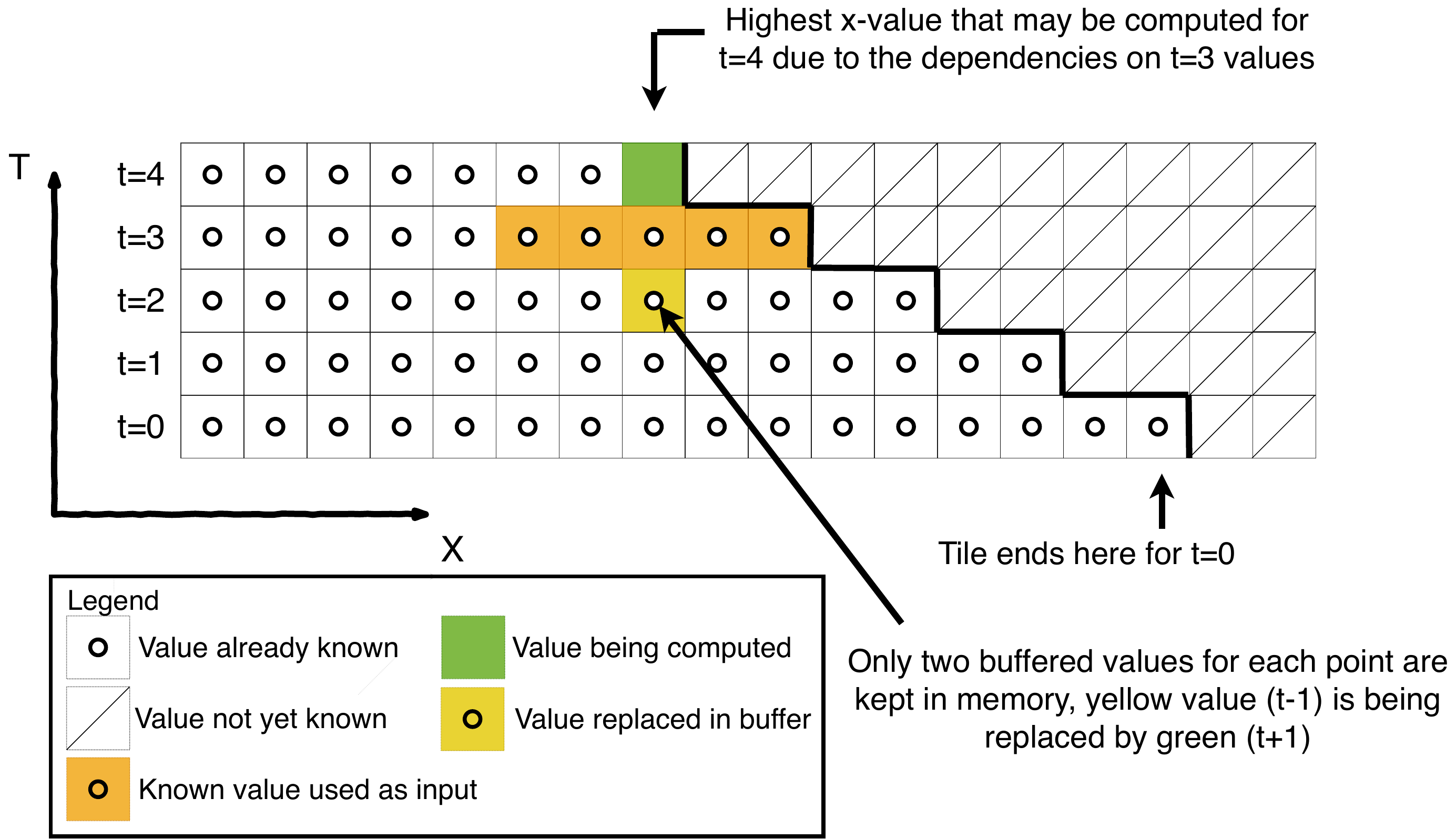}
	\caption{Illustration of stencil kernel update in WTB. The green point is updated using the orange values. This stencil kernel has a space order of 4, thus allowing a margin of 2 points on the right in order not to violate data dependencies. Only two timesteps are kept in memory (for time order one problems), so the green value substitutes the yellow one. Figure partially adapted from \cite{Yount2016}.}
	\label{fig:txy_wavefront}
\end{figure}

\begin{figure}[!ht]
	\centering
	\begin{subfigure}[b]{0.48\textwidth}
		\centering
		\includegraphics[width=0.90\textwidth]{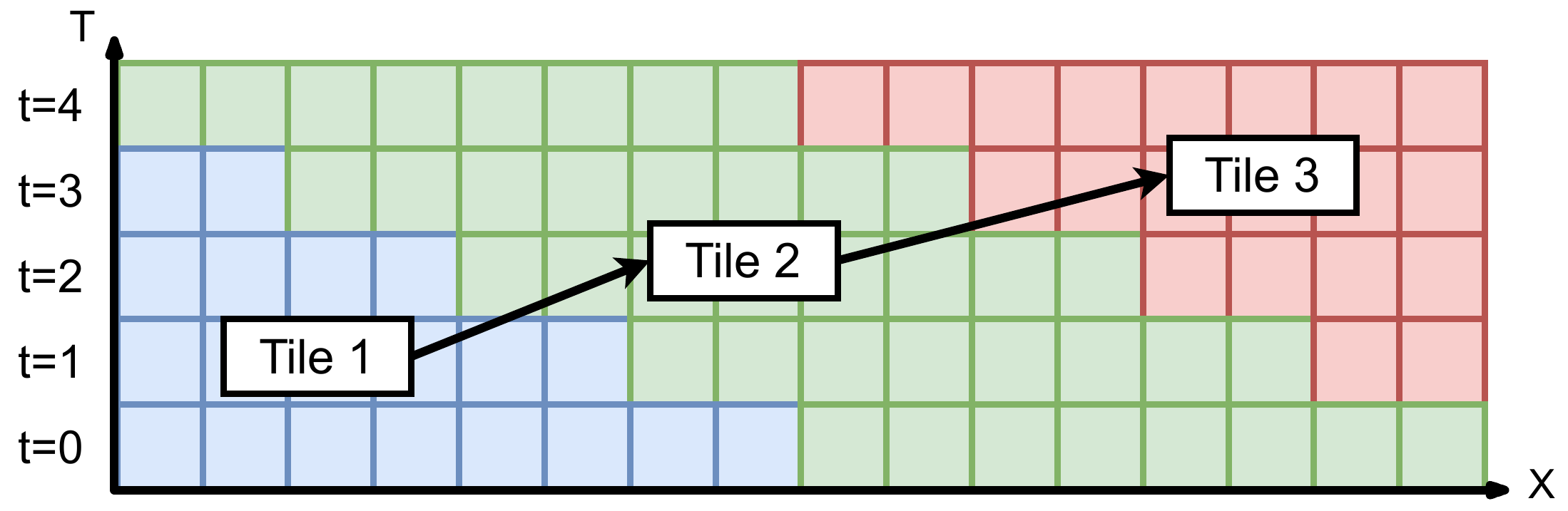}
		\caption{The figure shows multiple wave-front tiles evaluated sequentially, partially adapted from \cite{Yount2016}.}
		\label{fig:seq_wavefront}
	\end{subfigure}
	\hfill
	\begin{subfigure}[b]{0.48\textwidth}
		\centering
		\centering
		\includegraphics[width=0.90\textwidth]{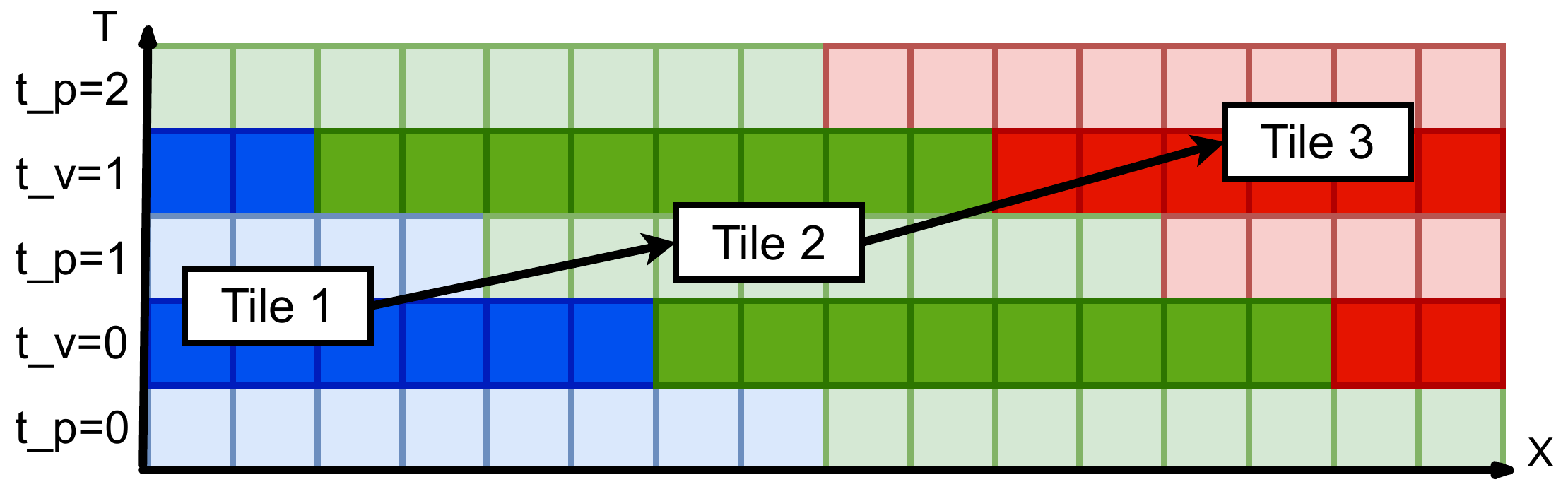}
		\caption{The figure shows multiple wave-front tiles evaluated sequentially in multigrid stencil codes.}
		\label{fig:stag_seq_wavefront}
	\end{subfigure}
	\hfill
    \caption{Wave-front updates for single- and multi-grid stencil updates.}
    \label{fig:wavefronts}
\end{figure}

After precomputing source injection, data dependencies are now aligned with the computational grid points. Applying temporal blocking is now feasible. We split the time-space iteration space into tiles as shown in Figure \ref{fig:seq_wavefront}. Each tile is then partitioned into space blocks. By applying the transformations required from wave-front temporal blocking to Listing \ref{lst:new_struct} we now have the loop structure \ref{lst:final_struct}. This structure is a time-tiled wave-front scheme over a stencil kernel update with source injection.

\begin{algo2lst}[!htbp]
	\For{ {\tt t\_tile} \textbf{in} {\tt time\_tiles}}{
		\For{{\tt  xtile} \textbf{in}  {\tt xtiles}}{
			\For{{\tt  ytile} \textbf{in} {\tt ytiles}}{
				\For{{\tt t} \textbf{in} {\tt t\_tile}}{
				\textbf{OpenMP parallelism}\\
				\For{{\tt xblk} \textbf{in} {\tt xtile}}{
						\For{{\tt yblk} \textbf{in} {\tt ytile}}{
							\For{{\tt x} \textbf{in} {\tt xblk}}{
								\For{{\tt y} \textbf{in} {\tt yblk}}{
									\textbf{SIMD vectorization}\\
									\For{ z = {\tt 1} \textbf{to} {\tt nz}}{
										$A(t, x-time, y-time, z)$;
										}
										\For{ z2 = {\tt 1} \textbf{to} {\tt nnz\_mask[x][y]}}{
											$I(t, x-time, y-time, z2)$;
										}
								}
							}
						}
					}
				}
			}
		}
	}
	\caption{The figure shows the loop structure after applying our proposed scheme.}\label{lst:final_struct}
\end{algo2lst}

The next section provides details about the evaluated kernels, their data dependencies, and their inherent loop structure.

\section{Structure of wave-propagation kernels}\label{kernels}

To illustrate our technique, we selected three representative kernels implementing explicit FD methods for wave propagation. The chosen kernels significantly differ in the operational intensity and working set size \cite{Louboutin2017a}. The kernels are implemented and validated in the \devito framework. The \devito compiler generates a C implementation for each kernel given a symbolic specification expressed with the \devito DSL.

\subsection{Isotropic acoustic}

The first equation we consider is the most straightforward and generally known wave-equation in an anisotropic acoustic medium. This equation is a single scalar PDE with a Jacobi-like stencil. The acoustic wave equation for the square slowness $m$, defined as $m=\frac{1}{c^2}$, where $c$ is the speed of sound in the given physical media, and a source $q$ is given by:

\begin{equation}
    \begin{cases}
     m(x) \frac{\partial^2 u(t, x)}{\partial t^2} - \Delta u(t, x) = \delta(x_s) q(t) \\
     u(0, .) = \frac{\partial u(t, x)}{\partial t}(0, .) = 0 \\
     d(t, .) = u(t, x_r).
     \end{cases}
    \label{acou}
\end{equation}

where $u(t, x)$ is the pressure wavefield, $x_s$ is the point source position, 
$q(t)$ is the source time signature, $d(t, .)$ is the measured data at positions $x_r$ and $m(x)$ is the
squared slowness. This equation writes in few lines with the \devito symbolic API as follows:

\begin{python}[label=WE, caption=Wave-equation symbolic definition]
from devito import solve, Eq, Operator
eq = m * u.dt2 - u.laplace
update = Eq(u.forward, solve(eq, u.forward))
src_eqns = s.inject(u.forward, expr=s*dt**2/m)
d_eqns = d.interpolate(u)
\end{python}

The discretized acoustic wave-equation is generally memory-bound due to the low computational count of the standard Laplacian \cite{Louboutin2017a, Williams2009}.

\subsection{Anistropic acoustic}

The second wave-equation kernel we consider is the most commonly used in industrial applications for subsurface imaging (RTM, FWI) \cite{zhang2011stable, louboutin2018effects, duveneck, Alkhalifah2000AnAW, Bube2016self}. This equation is a pseudo-acoustic anisotropic equation that consists of a coupled system of two scalar PDEs. Unlike the most simple acoustic isotropic equation, this formulation considers direction-dependent propagation speeds that translate into the discretized equation into a rotated anisotropic laplacian. Such a kernel increases the operation count drastically \cite{Louboutin2017a}. For example, the first dimension $x$ component of the Laplacian is defined as:

\begin{equation}
    \begin{aligned}
      G_{\bar{x}\bar{x}} &= D_{\bar{x}}^T D_{\bar{x}} \\
      D_{\bar{x}} &= \cos(\mathbf{\theta})\cos(\mathbf{\phi})\frac{\partial}{\partial x} + \cos(\mathbf{\theta})\sin(\mathbf{\phi})\frac{\partial}{\partial y} - \sin(\mathbf{\theta})\frac{\partial}{\partial z}.
    \end{aligned}
\label{rot}
\end{equation}

where $\mathbf{\theta}$ is the (spatially dependent) tilt angle (rotation around $z$), $\mathbf{\phi}$ is the (spatially dependent) azimuth angle (rotation around $y$). A more detailed description of the physics and discretization can be found in \cite{zhang2011stable, louboutin2018effects}.

\subsection{Isotropic elastic}

Finally, we consider the isotropic elastic equation. Unlike the two previous acoustic approximations, this equation has two significant properties. First, this is a first-order system in time, which allows us to extend our work to a smaller range of local data dependency over time. Consequently, we demonstrate that the benefits of time-blocking and our implementation of it are not limited to a single pattern along the time dimension. Second, this equation is a coupled system of a vectorial and a tensorial PDE, which increases the data movement drastically (one or two versus nine state parameters) on the wavefield and contains non-scalar expressions of the source and receiver expressions that involve multiple wavefields.

The isotropic elastic wave-equation, parametrized by the Lamé parameters $\lambda, \mu$ and the density $\rho$, is defined as \cite{virieux1986p}:
\begin{equation}
\begin{aligned}
&\frac{1}{\rho}\frac{\partial v}{\partial t} = \nabla . \tau \\
&\frac{\partial \tau}{\partial t} = \lambda \mathrm{tr}(\nabla v) \mathbf{I}  + \mu (\nabla v + (\nabla v)^T)
\end{aligned}
\label{elas1}
\end{equation}
where $v$ is a vector-valued function with one component per cartesian direction, and the stress $\tau$ is a symmetric second-order tensor-valued function.

In the following section, we consider these three wave-equations for varying spatial discretization orders to verify and analyze our temporal blocking method.

\section{Experimental Evaluation}\label{evaluation}

We outline in subsection \ref{setup} the experimental setup followed for performance evaluation. We aim to demonstrate the performance improvement achieved by our approach, illustrate its potential for impact on key applications, and probe its applicability limits.

\subsection{Compiler and system setup}\label{setup}

To evaluate our scheme, we used Virtual Machines in Azure with two architectures: Intel® Xeon® Processor E5 v4 Family (formerly called Broadwell) and  Intel® Xeon® Scalable Processors (formerly called Skylake 8171M). For our experiments, access was granted on VMs (on Microsoft Azure) called Standard\_E16s\_v3 and Standard\_E32s\_v3, running Ubuntu 18.04.4. The first system, E16s\_v3, has a single socket 8-core Intel Broadwell E5-2673 v4 CPUs with AVX2 support. Each Intel Broadwell CPU has three cache levels: L1 (32KB) and L2 (256KB) caches private to each core and a 50MB L3 cache shared per socket. The second system has single-socket 16-core IntelSkylake Platinum 8171M CPUs with AVX512 support. Each Intel Skylake CPU has three cache levels: L1 (32KB) and L2 (1MB) caches private to each core and a 35.75MB L3 cache shared per socket. Compilers used were GCC 7.5.0* and ICC 2021.1. We used OpenMP shared-memory parallelism with dynamic scheduling and SIMD vectorization. Thread pinning was enabled using the environment variables OMP\_PROC\_BIND (for GCC) and KMP\_AFFINITY (for ICC). Experiments were built with Devito v.4.2.3. The experimentation framework and instructions on reproducibility are available in subsection \ref{code}.
 
\subsection{Test case setup}\label{testcase}

We evaluate the performance of operators relevant to seismic imaging. We model the propagation of waves for three different models: \emph{ isotropic acoustic}, \emph{ anisotropic acoustic (TTI)} and \emph{ isotropic elastic}. The isotropic acoustic and TTI wave equations are discretized with second order in time while isotropic elastic with first-order, and we study varying space orders of 4, 8, 12. For all test cases, we use zero initial conditions and damping fields with absorbing boundary layers. Waves are injected using one time-dependent, spatially localized seismic source wavelet into the subsurface model. We benchmark velocity models of $512^3$ grid points, with a grid spacing of 10 for isotropic and elastic and 20 for TTI. Wave propagation is modeled in single precision for 512ms, resulting in 228 time-steps for isotropic acoustic, 436 for isotropic elastic, and 587 for anisotropic acoustic. The time-stepping interval is selected regarding the Courant-Friedrichs-Lewy (CFL) condition \cite{Courant1967}, ensuring the explicit time-stepping scheme's stability determined by the highest velocity of the subsurface model and the grid spacing.

\begin{figure*}[!htbp]
	\centering
	\begin{subfigure}[b]{0.45\textwidth}
		\includegraphics[width=\textwidth]{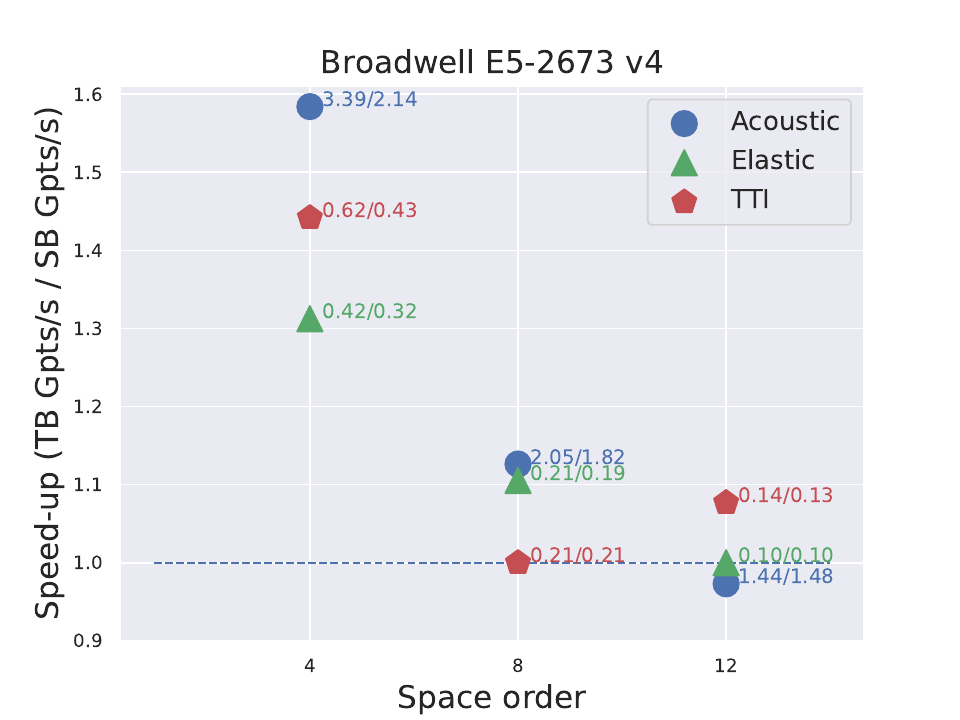}
		\caption{Throughput speed-up of kernels for Broadwell.}
		\label{fig:BDWspeedup}
	\end{subfigure}
	\begin{subfigure}[b]{0.45\textwidth}
		\includegraphics[width=\textwidth]{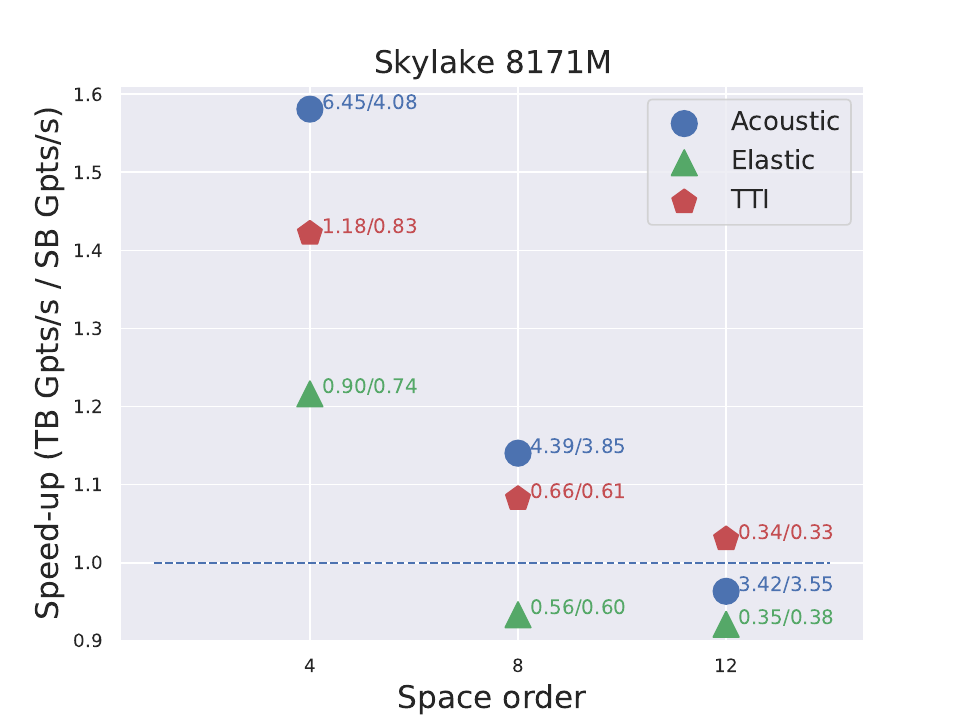}
		\caption{Throughput speed-up of kernels for Skylake.}
		\label{fig:SKLspeedup}
	\end{subfigure}
	\caption{Throughput speed-up of temporal blocking kernels versus highly-optimised vectorized spatially-blocked code in Intel Xeon architectures, Broadwell and Skylake.}
	\label{fig:speedup}
\end{figure*}

\subsection{Autotuning temporally blocked code}

It should be noted that the parameter space for temporal blocking schemes is extensive. We report results obtained from guidance and experience from the state-of-the-art codes and literature \cite{Wittman2010}. Simulation codes are hard to generalize in terms of performance as multiple configurations may be used from case to case. An operator's performance depends upon many factors, such as grid shape, discretization space order, tile and block shapes, number of other fields, number of timesteps, platforms, and others. To tune our C code for the underlying hardware, we swept over the whole parameter space to find the global performance maxima. We executed our experiments using the best-performing tile and block sizes, ensuring a fair comparison versus \devito 's aggressively tuned optimized spatially-blocked and vectorized code. The best performing tile sizes for temporal blocking are reported in Table \ref{tile_block_sizes}. Figure \ref{fig:speedup} illustrates the throughput (GPoints/s) speedup achieved for each evaluated model for space order discretizations of 4, 8, and 12.

\begin{table}[!htbp]
	\centering
	\begin{tabular}{m{2cm} m{2.5cm} m{2.5cm}}
		\toprule
		        & ${tile_x,tile_y,block_x,block_y}$ \tabularnewline
		Problem       & \bf{Broadwell}& \bf{Skylake} \tabularnewline
		\midrule
		Acoustic O(2,4)    & 32, 32, 8, 8   & 64, 64, 8, 8   \tabularnewline
		Acoustic O(2,8)    & 64, 64, 8, 8   & 64, 64, 8, 8   \tabularnewline
		Acoustic O(2,12)   & 256, 256, 8, 8 & 128, 128, 8, 8 \tabularnewline
		Elastic O(1,4)     & 32, 32, 8, 8   & 32, 32, 8, 8   \tabularnewline
		Elastic O(1,8)     & 32, 32, 8, 8   & 64, 56, 8, 12  \tabularnewline
		Elastic O(1,12)    & 256, 256, 8, 8 & 256, 256, 8, 8 \tabularnewline
		TTI O(2,4)         & 40, 32, 4, 4   & 48, 48, 8, 8   \tabularnewline
		TTI O(2,8)         & 32, 32, 8, 8   & 64, 64, 8, 8   \tabularnewline
		TTI O(2,12)        & 256, 256, 8, 8 & 256, 256, 8, 8  \tabularnewline

		\bottomrule
	\end{tabular}
	\caption{Optimal tile-block shapes after tuning WTB}\label{tile_block_sizes}
\end{table}

\subsection{Results discussion}

Figure \ref{fig:speedup} illustrates the speedup achieved for each of our wave propagation models. All of our models show speedup for space order four discretization on both platforms. Acoustic benefits the most with around 1.6x and TTI follows with around 1.44x. Elastic wave propagation is accelerated by 1.3x on Broadwell and 1.22x on Skylake. Concerning space order 8, a commonly used practice, we observe speedups of 1.13x or more for acoustic, elastic on Broadwell and acoustic, TTI on Skylake. No significant performance gains are observed for space order 12, excluding some gains of around 5\% on Broadwell with isotropic elastic and TTI. Figure \ref{fig:roofline} shows the isotropic acoustic kernels' roofline performance for the Broadwell microarchitecture. The roofline is a cache-aware roofline model representing cumulative (L1+L2+LLC+DRAM) traffic-based Arithmetic Intensity for application kernels \footnote{https://software.intel.com/content/www/us/en/develop/articles/integrated-roofline-model-with-intel-advisor.html}. We showcase improvement for the acoustic model breaking the ceiling of the L3 cache.

\subsection{Corner cases}\label{corner_cases}

Although our test cases use a single source, it is interesting to explore how our model performs with the presence of more off-the-grid operators. Each source is decomposed into its surrounding grid points, so the overhead increases due to the number of initial sources and the number of grid points affected. We evaluate the overhead induced for two cases a) an increasing number of sparsely located sources: in this case, we have an increasing number of sources located at an x-y plane slice of the 3D grid, a scenario which can be of practical interest and b) an increasing number of sources densely and uniformly located all over the 3D grid. Figure \ref{fig:dense} shows that for isotropic acoustic wave propagation, the increasing number of sources is not affecting performance gains except with really densely located sources where our scheme is not taking advantage of the structure sparsity. Still, though, we observe gains of around 1.4x compared to 1.55x previously.

\begin{figure}[!htpb]
	\centering
	\includegraphics[width=0.45\textwidth]{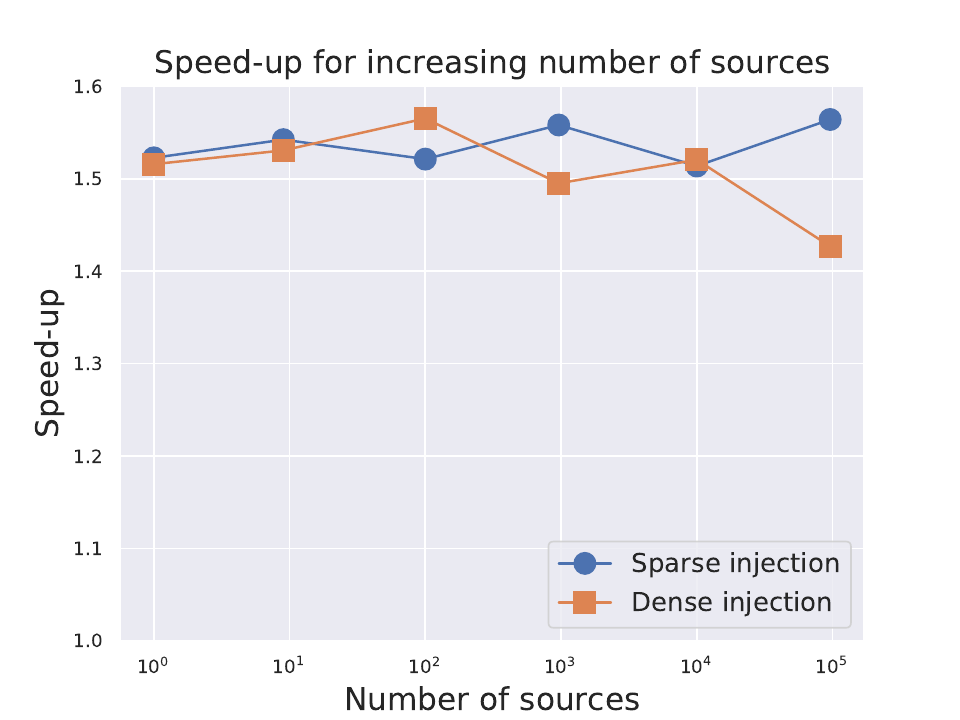}
	\caption{Throughput speed-up for an isotropic acoustic operator for space order 4 over an increasing number of sources, sparsely and densely located.}
	\label{fig:dense}
\end{figure}

\begin{figure}[!htpb]
	\centering
	\includegraphics[width=0.48\textwidth]{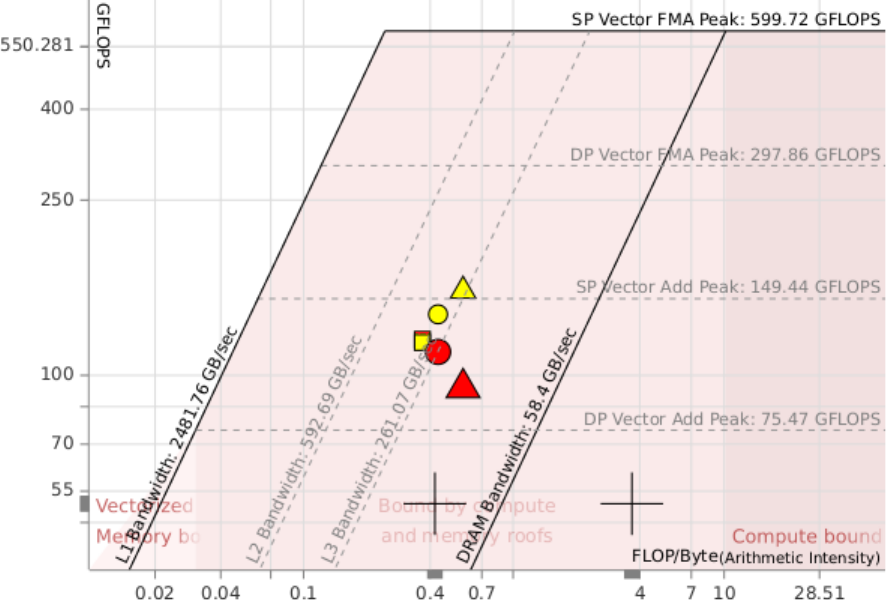}
	\caption{Cache-aware roofline model on Broadwell for isotropic acoustic model space order 4 (triangles), 8 (circles), and 12 (squares). Red markers show the performance of spatially blocked vectorized kernels, while yellow ones show our temporal blocking scheme's performance.}
	\label{fig:roofline}
\end{figure}

\section{Conclusions}\label{conclusion}
This paper introduced a mechanism to enable temporal blocking in stencil computations involving sparse off-the-grid operators as encountered, for example, with sources and receivers in seismic inversion problems. We applied wave-front temporal blocking to wave-propagators ranging from isotropic acoustic to more advanced, such as isotropic elastic and anisotropic acoustic (TTI). Experimental evaluation of the improved kernels on Broadwell and Skylake microarchitectures showed compelling evidence of substantial acceleration of at least 1.5x for low and at least 1.1x for medium space order wave-propagation kernels.

\subsection{Code availability}\label{code}

An implementation of the methods described in this paper is available in a \devito fork repository under the MIT open-source license at \href{https://github.com/georgebisbas/devito/releases/tag/v0.9-alpha}{georgebisbas/devito v0.9-alpha}. See the \href{https://github.com/georgebisbas/devito/blob/v0.9-alpha/README.md}{README.md} for instructions on how to reproduce the results in the paper.

\subsection{Future work}

Achieving performance improvement with high-space order kernels requires further research work. Methods like stencil retiming \cite{Stock} have shown promise in alleviating this performance bottleneck, and a possible combination with temporal blocking may be promising. Another possible solution can be data layout transformations \cite{Yount2015}. Near-term plans include evaluating our scheme on more diverse architectures (e.g., ARM) and accelerators (e.g., GPUs). The next step is the full automation and integration in the \devito DSL \cite{LuporiniTOMS}. We aim to deliver automated, scalable optimizations on generated code beyond our kernels' current roofline performance limit. This paper's evaluation results are mainly motivated by the seismic imaging domain; however, the target applications are not limited to this scope.

\section*{Acknowledgments}
This research is funded by the Engineering and Physical Sciences Research Council (EPSRC) grants  EP/L016796/1, EP/R029423/1, EP/V001493/1, and HiPEDS Center for Doctoral Training. The author thanks Richard M. Veras, Navjot Kukreja, John Washbourne, and Giacomo La Scala for the fruitful discussions as well as the whole \devito community.

\bibliography{sparsebib}


\end{document}